\documentclass[11pt,a4paper]{article}
\usepackage{jheppub_kim}
\usepackage{rotating} 
\usepackage{graphicx,epsfig}
\usepackage{amsmath}
\usepackage {amssymb}
\usepackage{subfigure}

\usepackage{relsize}

\providecommand{\U}[1]{\protect\rule{.1in}{.1in}}

\newcommand{\newc}{\newcommand}

\newc{\be}{\begin{equation}}
\newc{\ee}{\end{equation}}

\newc{\ba}{\begin{eqnarray}}
\newc{\ea}{\end{eqnarray}}
\newc{\bea}{\begin{eqnarray*}}
\newc{\eea}{\end{eqnarray*}}
\newc{\D}{\partial}
\newc{\ie}{{\it i.e.} }
\newc{\eg}{{\it e.g.} }
\newc{\etc}{{\it etc.} }
\newc{\etal}{{\it et al.}}
\newc{\lcdm}{$\Lambda$CDM }

\newc{\ra}{\Rightarrow}

\title{Inflation from a generalized exponential plateau: towards  extra 
suppressed 
tensor-to-scalar ratios}

\author[a,b]{Gerasimos Kouniatalis}

\author[b,c,d]{Emmanuel N. Saridakis}

\affiliation[a]{Physics Department, National Technical University of Athens,
15780 Zografou Campus,  Athens, Greece}
\affiliation[b]{National Observatory of Athens, Lofos Nymfon, 11852 Athens, 
Greece}

\affiliation[c]{Department of Astronomy, School of Physical Sciences, 
University of Science and Technology of China, Hefei 230026, P.R. China}

 \affiliation[d]{Departamento de Matem\'{a}ticas, Universidad Cat\'{o}lica del 
Norte, 
Avda.
Angamos 0610, Casilla 1280 Antofagasta, Chile}

\emailAdd{gkouniatalis@noa.gr}
 \emailAdd{msaridak@noa.gr}

\abstract{ 
 We investigate a standard minimally-coupled scalar-field 
inflationary scenario,  which is based on a new  potential, 
with suitably generalized plateau features, that 
leads to  extra small tensor-to-scalar ratios. In particular, we consider a 
specific three-parameter potential, which has a flatter plateau and a 
steeper well compared to the Starobinsky potential in the Einstein frame. 
We study  the inflationary realization and we show   that it guarantees a 
prolonged period of slow-roll inflation and a successful exit. Additionally, 
the 
steeper minimum   leads to significantly 
suppressed tensor perturbations, and thus to an extra-small tensor-to-scalar 
ratio $r$, and  we show that we are able to obtain $r$ values  less than 
$10^{-5}$.  Moreover, we calculate the reheating temperature showing 
that in order to be in agreement with observations one of the potential 
parameters should remain within specific bounds.
Finally, performing an inverse conformal transformation to the Jordan frame 
we show  that the considered potential corresponds to higher-order corrections 
to Starobinsky potential in the Einstein frame, and these corrections are the 
reason for the improved behavior of the tensor-to-scalar ratio.

}

\begin{document}
\maketitle

\section{Introduction}

Inflation has become an essential component of the Standard Model of Cosmology
\cite{Starobinsky:1980te,Kazanas:1980tx,Sato:1980yn,Guth:1980zm,Linde:1981mu}.
It offers successful solutions to the horizon and flatness problems and yields 
robust predictions, such as a nearly scale-invariant spectrum of primordial 
perturbations, all of which have been supported by observations of the cosmic 
microwave background (CMB) radiation.
Despite its empirical success, the exact underlying mechanism responsible for 
producing the inflationary phase and the successful exit from it, remains one 
of the unresolved questions in both modern particle physics and 
cosmology. Hence, the theoretical effort to construct models capable of 
explaining this early period of accelerated cosmic expansion has increased 
  in recent years. In general, these models fall into two principal 
categories: those arising from the dynamics of a scalar field, and those 
emerging 
from modifications to gravity itself (for reviews, see
\cite{Olive:1989nu, 
Lyth:1998xn,Bartolo:2004if,Nojiri:2010wj,Martin:2013tda}).

A successful inflationary model must yield predictions consistent with detailed 
cosmological observables extracted from measurements of the CMB anisotropies 
and large-scale structure. The most fundamental 
observables include the amplitude of the scalar power spectrum $A_s$, the 
spectral index of scalar perturbations $n_s$, the tensor-to-scalar ratio $r$, 
and
the running of the scalar spectral index $\alpha_s \equiv 
\mathrm{d}n_s/\mathrm{d}\ln k$, and moreover one should be consistent with  
non-Gaussianities, too \cite{Martin:2013tda}. The amplitude $A_s$ is fixed by 
observations to be of 
order $10^{-9}$, while the spectral index $n_s$ is tightly constrained around 
$n_s \approx 0.965$, indicating a slight deviation from scale invariance. 
Therefore, any theoretically consistent inflationary scenario must not only 
reproduce a sufficiently long phase of accelerated expansion but also yield a 
scalar perturbation spectrum and a suppressed tensor contribution in agreement 
with current data.

Perhaps most constraining, however, is the tensor-to-scalar ratio $r$, which 
quantifies the relative contribution of primordial gravitational waves to the 
CMB. Current observations from Planck and BICEP/Keck Array 
\cite{Planck:2018jri,BICEP:2021xfz} place a stringent upper bound $r < 0.036$ 
(at 95\% confidence level). Hence, this significant observable by itself was 
capable of excluding the traditional  large-field models based on simple 
monomial potentials, since they typically predict  significant tensor 
amplitudes, and opened the research towards extended inflationary models with 
plateau-like potentials, obtained from various non-minimal coupled 
scalar-field models
 \cite{Faraoni:1996rf,Bastero-Gil:2006zpr,Nojiri:2007bt,Bamba:2008ja,
Einhorn:2009bh,Bauer:2008zj,Germani:2010gm,Feng:2010ya,Hertzberg:2010dc,
Hossain:2014xha,Hossain:2014coa,Hossain:2014ova,Geng:2015fla,Lola:2020lvk, 
Braglia:2020fms,
Karydas:2021wmx, Papanikolaou:2022did,Sohail:2024oki}, 
Horndeski  
\cite{Tsujikawa:2014mba,Germani:2015plv,BeltranJimenez:2017cbn,
Sebastiani:2017cey,Oikonomou:2020sij,Chen:2021nio}
 and Galileon   
\cite{Kobayashi:2010cm,Burrage:2010cu,Renaux-Petel:2011lur,Renaux-Petel:2011rmu,
Choudhury:2012whm,Ohashi:2012wf,Qiu:2015nha,Choudhury:2023hvf,
Choudhury:2023hfm} 
theories, etc. Alternatively, a large number of 
inflationary 
models has been constructed in the framework of modified gravity  
\cite{CANTATA:2021ktz,Barrow:1988xh,Garcia-Bellido:1995him,Nojiri:2003ft,
Carter:2005fu,Ferraro:2006jd,Nojiri:2007as,Nojiri:2007cq, 
Mukohyama:2009gg,Nozari:2010kri,Elizalde:2010ep,Cai:2010kp,Briscese:2012ys,
Sebastiani:2013eqa,Bamba:2014jia,Bamba:2014mua,Nojiri:2014zqa,Nojiri:2015fra,
DeLaurentis:2015fea,Bamba:2016wjm,Geng:2017mic,Paliathanasis:2017apr,
Nojiri:2017ncd,Awad:2017ign,
Chakraborty:2018scm, Qiu:2018nle, 
Shimada:2018lnm,Benisty:2019jqz,Gamonal:2020itt,Cai:2021uup, 
Papanikolaou:2021uhe,Shiravand:2022ccb,Panda:2022can,Bhat:2023qwa,
Sadatian:2024lub,Basilakos:2023xof, Tzerefos:2023mpe, Papanikolaou:2024cwr}, 
with 
the notable example being  Starobinsky $R^2$ inflation 
\cite{Starobinsky:1980te}, which can lead to a 
very small $r$ (of the order of $\sim$0.003-0.005).

The fact that the likelihood contours for the tensor-to-scalar ratio $r$ always 
include the zero value, led to the construction of inflationary models that can 
lead to even smaller $r$ values. One class of such models are the 
$\alpha$-attractors 
\cite{Kallosh:2013hoa,Kallosh:2013yoa,Kallosh:2013tua,Galante:2014ifa,
Linde:2015uga, 
Gao:2017uja,Miranda:2017juz,Dimopoulos:2017zvq,Dimopoulos:2017tud, 
Scalisi:2018eaz, Braglia:2020bym, Bhattacharya:2022akq,  
Brissenden:2023yko,Ferrara:2016fwe,Akarsu:2016qhf}. These 
models are characterized by a non-canonical kinetic term  which leads to a pole 
in the kinetic function, and thus upon canonical normalization, this kinetic 
structure gives rise to a universal plateau-like potential for large field 
values, such that the inflationary predictions become largely independent from 
the detailed shape of the potential. 
 In the limit $\alpha \to 0$, the models tend asymptotically to the 
predictions of Starobinsky inflation, while larger values of $\alpha$ 
interpolate toward chaotic inflation. 

In this work we are interested in suggesting a new potential, with suitably 
generalized plateau features that are able to lead to extra small 
tensor-to-scalar ratios without the need to consider non-minimal extensions of 
scalar-field cosmology. In particular, we consider a specific three-parameter 
potential which has a flatter plateau and a steeper well comparing to 
Starobinsky potential in the Einstein frame, which leads to significantly 
smaller $r$ values. Transforming it to the Jordan frame we can see that it 
corresponds to higher-order extensions to Starobinsky  $R^2$ gravity, and these 
extensions are the cause of the improved behavior.

 The plan of the work is the following: In Section \ref{model} we present 
the potential with generalized exponential plateau, and we investigate in 
detail the  
obtained inflationary dynamics. Additionally, we calculate the reheating 
temperature and we examine the conformal transformation to the Jordan frame 
and the relation to gravitational modification. Finally, the conclusions 
summarize the key findings of our work.

\section{Inflationary dynamics from    generalized exponential plateaus}
\label{model}

In this section we present the explicit inflation model, based on a new  
potential with generalized exponential plateau, and then we investigate in 
detail the inflationary dynamics. First we present the new model and then we 
examine in detail the inflationary realization.

\subsection{Potentials with  generalized exponential plateaus}

We consider the action 
\begin{equation}\label{action}
\mathcal{S}=\int d^{4}x \sqrt{-g} \left[  \frac{M^{2}_{Pl}}{2}R -\frac{1}{2} 
g^{\mu\nu}\partial_{\mu} \phi \partial_{\nu} \phi-V(\phi) \right],
\end{equation}
with   $M_{Pl}\approx 2.4 \times 10^{18}GeV$   the reduced Planck mass, 
which describes a scalar field $\phi$ minimally coupled to gravity. For the 
potential we consider the form
\begin{equation}\label{potential}
    V(\phi) = V_0 \left( 1-e^{\frac{-\alpha \phi^2}{\beta M_{Pl}^2 +\gamma 
\phi^2}}  
\right),
\end{equation}
where $V_0$, { and the dimensionless $\alpha$, $\beta  $, $\gamma$,} are 
the model parameters.

{The motivation behind the above potential is to construct an analytic 
potential that simultaneously has (i) a finite large-field plateau of 
tunable height, (ii) an approximately quadratic minimum to ensure standard 
reheating via coherent oscillations, and (iii) a single crossover scale 
producing a prolonged but controlled slow roll. These three requirements point 
naturally to an exponential of rational functions of $\phi^2$ of the form
(\ref{potential}), which indeed yields a quadratic expansion near the origin. 
Moreover, as we will see in the following, the above potential does have  a 
very flat finite plateau as $\phi\!\to\!\infty$, which leads to very small 
slow-roll parameters and thus to extra 
suppressed tensor-to-scalar ratios. Finally, as we will also discuss in the 
following, under inverse conformal transformation the 
Einstein-frame potential (\ref{potential}) corresponds to higher-order 
($R^n$-type) corrections to Starobinsky quadratic $f(R)$ model, hence in a 
sense it can be viewed as a   way to flatten the plateau while keeping a 
steeper quadratic well. }

In summary, potential (\ref{potential}) exhibits a 
flatter plateau and a steeper well comparing to other 
potentials of the literature, it recovers a quadratic form in 
the small-field limit   and a plateau-like behavior in the large-field regime, 
and as we will see   not only reproduces the successful predictions of both 
chaotic and plateau inflationary scenarios, but also provides a natural 
graceful 
exit from inflation. Furthermore, the transition from inflation to reheating is 
improved by coherent oscillations around the quadratic minimum, in line with
the established picture of the post-inflationary universe.
However, note that an important advantage of the above model is that the field 
is minimally coupled to gravity, and thus no extra couplings or non-canonical 
kinetic terms are introduced. Thus, the interesting potential dynamics arise 
solely from the potential, and not from the complications of the theory. 
 
As usual, we consider a   flat   Friedmann-Robertson-Walker (FRW) 
  metric, namely
\begin{equation}
\label{metric}
ds^{2}=-dt^{2}+a^{2}(t)\delta_{ij}dx^{i}dx^{j}\,,
\end{equation}
where $a(t)$ is the scale factor. 
Varying the action   (\ref{action}) with respect to the metric 
yields the two Friedmann equations 
\begin{equation}
\label{Fried1}
3 M^{2}_{Pl} H^2=  \frac{\dot{\phi}^2}{2}  +V_0 \left( 
1-e^{\frac{-\alpha \phi^2}{\beta M_{Pl}^2 +\gamma \phi^2}}\right) 
\end{equation}
\begin{equation}
\label{Fried2}
-M^{2}_{Pl} \dot H=  \frac{\dot{\phi}^2}{2}  ,
\end{equation}
 where  $H=\frac{\dot a}{a}$ is the  Hubble parameter. Moreover, variation 
with respect to the scalar field  leads to the Klein-Gordon equation, namely
\begin{equation}
\ddot \phi +3H\dot \phi + V'(\phi) =0,
\label{phiequation}
\end{equation}
where we denoted $V'(\phi)\equiv \frac{dV}{d\phi}. $
We mention that we can re-write the above   equation as the usual 
conservation equation $\dot{\rho}_{\phi}+3H(\rho_\phi+p_\phi)$, under the 
identifications
\begin{eqnarray}
&&
\rho_{\phi} = \frac{\dot{\phi}^2}{2}+V 
\nonumber\\ \nonumber\\
&& 
p_{\phi} = \frac{\dot{\phi}^2}{2}-V .
\end{eqnarray}

 Potential (\ref{potential}) is the novelty of the present work. Its rational 
structure, governed by the $\alpha$, $\beta M_{Pl}^2$ and $\gamma$ parameters, 
exhibits  
a  smooth transition from a quadratic minimum at $\phi = 0$ to a flat plateau 
at large field values, which  ensures that slow-roll inflation occurs 
naturally. This form allows $\gamma$ to act as a phenomenological parameter 
that 
controls the transition between the quadratic, chaotic-inflation form, and the 
strong plateau, enhancing the model. Finally, the model allows for a 
successful reheating, with oscillations around the quadratic minimum. In the 
next subsection we investigate the inflationary realization in detail.
  
\subsection{Inflation realization}
  
 Let us now perform a quantitative analysis of inflationary evolution. Since we 
have a minimally-coupled scalar-field model, we introduce the standard    
slow-roll parameters, namely \cite{Martin:2013tda}
\begin{eqnarray}
    &&\varepsilon_V = \frac{M_{Pl}^2}{2}\left(\frac{V'}{V}\right)^2 \nonumber\\
    && 
\eta_V = M_{Pl}^2 \frac{V''}{V}\nonumber\\
    &&  \xi^2_V= M_{Pl}^4 
\left(\frac{V'V'''}{V^2}\right). \label{slowrollparameters}
\end{eqnarray}
 As one 
can see, for large-field values the above expressions give 
\begin{eqnarray}
        \varepsilon_V &\approx& \frac{2M_{Pl}^6}{\phi^6}\left[\frac{\alpha 
\beta e^{-\alpha/\gamma}}{\gamma^2(1-e^{-\alpha/\gamma})}\right]^2 \nonumber\\
    \eta_V &\approx& -\frac{6\alpha \beta 
e^{-\alpha/\gamma}}{\gamma^2(1-e^{-\alpha/\gamma})} \frac{M_{Pl}^4}{\phi^4} 
\nonumber\\
\xi_V^2 
&\approx&
\frac{48 M_{\mathrm{Pl}}^8}{\phi^{8}}
\left[\frac{\alpha \beta  
e^{-\alpha/\gamma}}{\gamma^2(1-e^{-\alpha/\gamma})}\right]^2 ,
\end{eqnarray}
which can become suitable small, ensuring the inflation beginning, since
the slow-roll conditions, $\varepsilon\ll 1$ and $|\eta_V| 
\ll 1$, are satisfied. Nevertheless, when one of the slow-roll parameters 
reaches unity, we have the inflation exit.

Using the exact values for the slow-roll parameters  
(\ref{slowrollparameters}), 
and examining when they become 1, we can calculate the field value at the end 
of inflation  $\phi_{end}$ as 
\begin{equation}
    \phi_{end} = \sqrt[6]{ 2M_{Pl}^6 \left[ \frac{\alpha\beta   
e^{-\alpha 
/\gamma}}{\gamma^2(1-e^{-\alpha/\gamma})}\right]^2}. 
\end{equation}
As usual, the total number of e-folds is given by
\begin{equation}
    N_* = \int_{\phi_{end}}^{\phi_*} \frac{d\phi}{M_{Pl}\sqrt{2\varepsilon_V}},
\end{equation}
where $\phi_*$ is the field value at the onset of the   inflationary phase. The 
$\phi_*$ value can be calculated exactly  numerically, however one can extract 
an approximate analytical expression in the large-field approximation as 
\begin{equation}
     \phi_*^4 = \frac{8N_* M_{Pl}^4\alpha \beta  e^{-\alpha/\gamma}}{\gamma^2 
(1-e^{-\alpha/\gamma})} + \phi_{end}^4.
\end{equation}
 Finally, the inflationary observables, such as   the scalar spectral index of 
the curvature 
perturbations $n_\mathrm{s}$, the running $\alpha_\mathrm{s} \equiv d 
n_\mathrm{s}/d 
\ln k$ of the spectral index 
$n_\mathrm{s}$, where $k$ is the absolute value of the wave number $\Vec{k}$,
the tensor spectral 
index $n_\mathrm{T}$ and the tensor-to-scalar ratio $r$,  can be calculated as 
\cite{Martin:2013tda}
\begin{eqnarray}
  n_\mathrm{s} &\approx& 1-6\varepsilon_V+2\eta_V  ,
\label{eps2222} \\
\alpha_\mathrm{s} &\approx& 16\varepsilon_V\eta_V-24\varepsilon_V^2-2\xi^2_V , 
\label{eps333}
\\
r &\approx&16\varepsilon_V ,
 \label{eps111}\\
 n_\mathrm{T} &\approx& -2\varepsilon_V .
\label{eps444}
\end{eqnarray}

\begin{figure}[ht]
  \centering
  \includegraphics[width=0.58\textwidth]{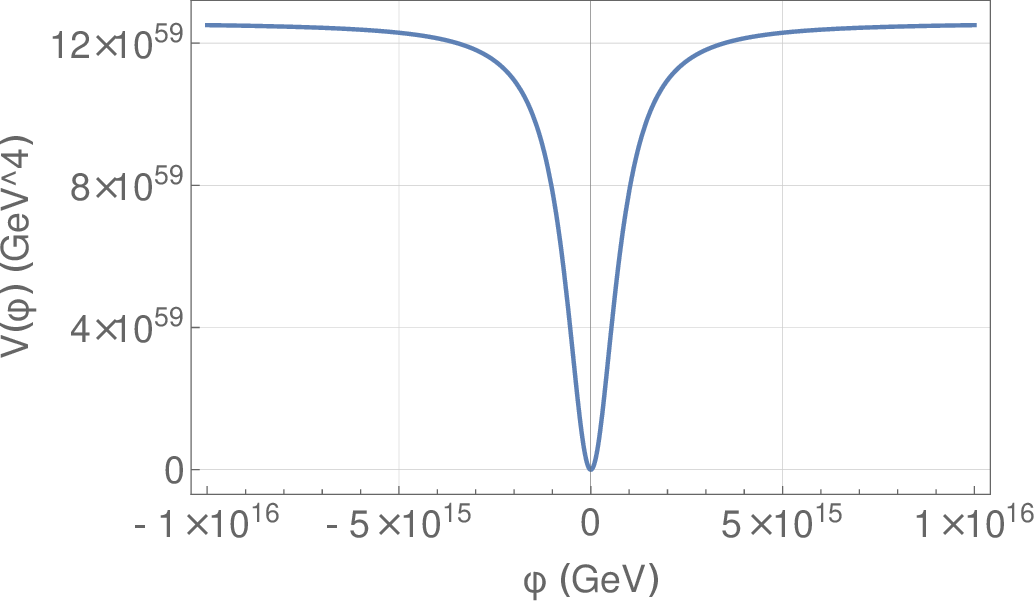}
  \caption{ {\it{The potential  (\ref{potential}) with generalized  exponential 
plateau proposed in this work,  for   $\alpha = \gamma = 1$, 
$\beta = 1.7\times 10^{-7}$, and $V_0=2  \times 10^{60}GeV^4$. }} 
  }
  \label{fig:potential}
\end{figure}

Let us proceed to the quantitative analysis. 
As we find, in order to satisfy the above value, and obtain  tensor-to-scalar 
ratio and scalar spectral index in agreement with their observed bounds, we 
need to choose $\gamma$ and $\alpha $ of the order of 1,  and 
$\beta = 1.7\times 10^{-7}$. For each choice, $V_0$ is adjusted in order for 
the 
scalar power spectrum 
amplitude, which quantifies the magnitude of density perturbations, to take its 
observed value \cite{Planck:2018jri}
\begin{equation}
    A_s \approx 2 \times 10^{-9}.
\end{equation} 
In Fig.  \ref{fig:potential}  we depict the potential  (\ref{potential}) for 
the above parameter choices.
 With these parameter choices, the model yields a viable inflationary 
trajectory: inflation begins when the inflaton field is approximately at 
$\phi_* 
\approx  1.9 \times 10^{17}\text{GeV}$
and terminates   at $\phi_{\text{end}}\approx 1.25 \times 10^{16}$GeV, after 
 $N_*=50$  e-foldings, while it 
 begins   at $\phi_* 
\approx 2 \times 10^{17}\text{GeV}$
and terminates   at $\phi_{\text{end}}\approx 1.3 \times 10^{16}$GeV, after 
 $N_*=60$  e-foldings.

 \begin{figure}[ht]
\centering
\hspace{-1.cm}
\includegraphics[scale=.48]{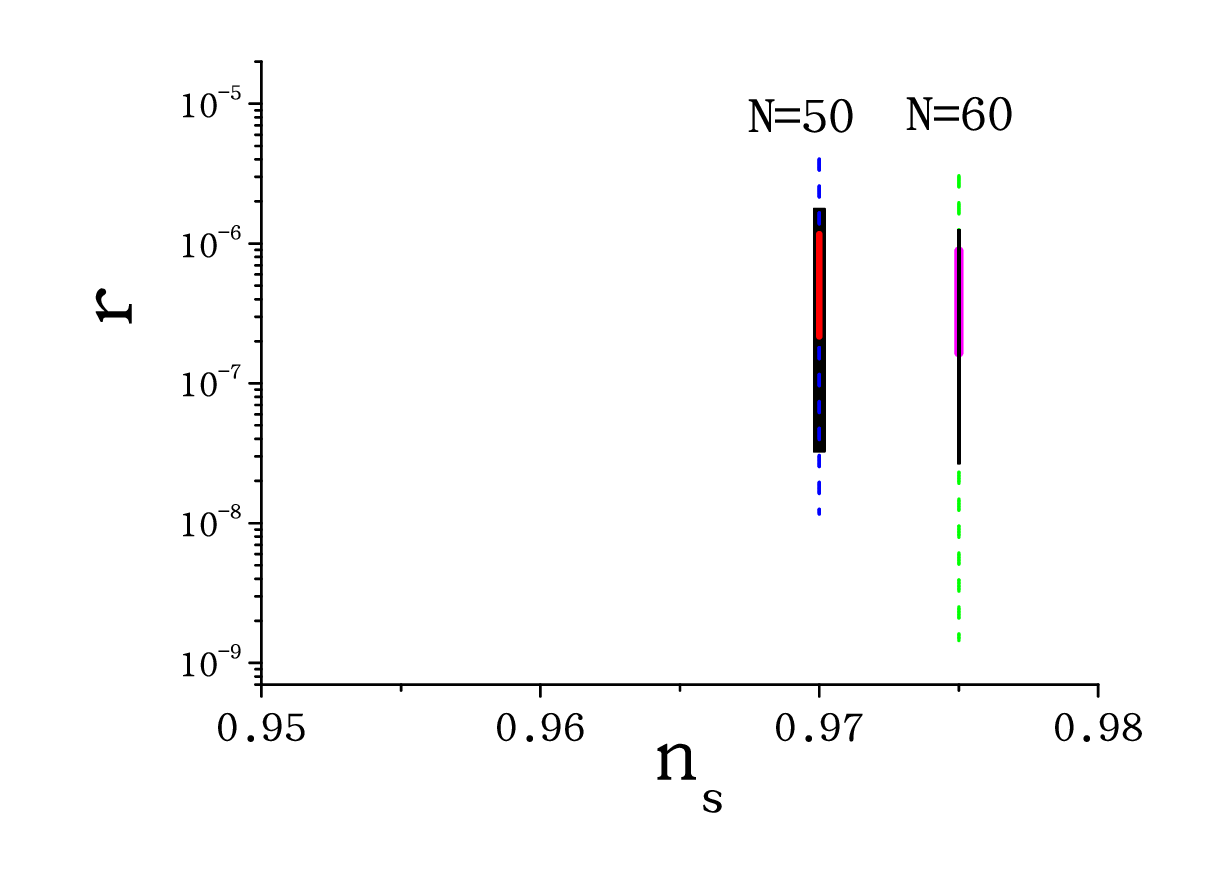}
\caption{{\it{ The predictions of the inflationary scenario based on the 
potential  (\ref{potential}) with generalized  exponential 
plateau on the  $n_{\mathrm{s}}-r$ plane. The left set of curves corresponds to 
e-folding number $N_*=50$, for $\gamma = 0.1 $ (blue 
dashed curve), $\gamma = 1 $ (black thick curve), and $\gamma = 2 $ 
(red curve). 
 The right set of curves corresponds to 
e-folding number $N_*=60$, for $\gamma = 0.1 $ (green 
dashed curve), $\gamma = 1 $ (black curve), and $\gamma = 2 $ 
(magenta curve). 
In all curves $\beta = 1.7\times 10^{-7}$,  $V_0$ is set around $V_0=2  
\times 10^{60}GeV^4$ in order to obtain   $A_s = 2 \times 10^{-9}$, and 
$\alpha$ runs from $0.1$ to $10$ .
}}}
\label{rns}
\end{figure}

\begin{figure}[ht]
\centering
\hspace{-1.cm}
\includegraphics[scale=.48]{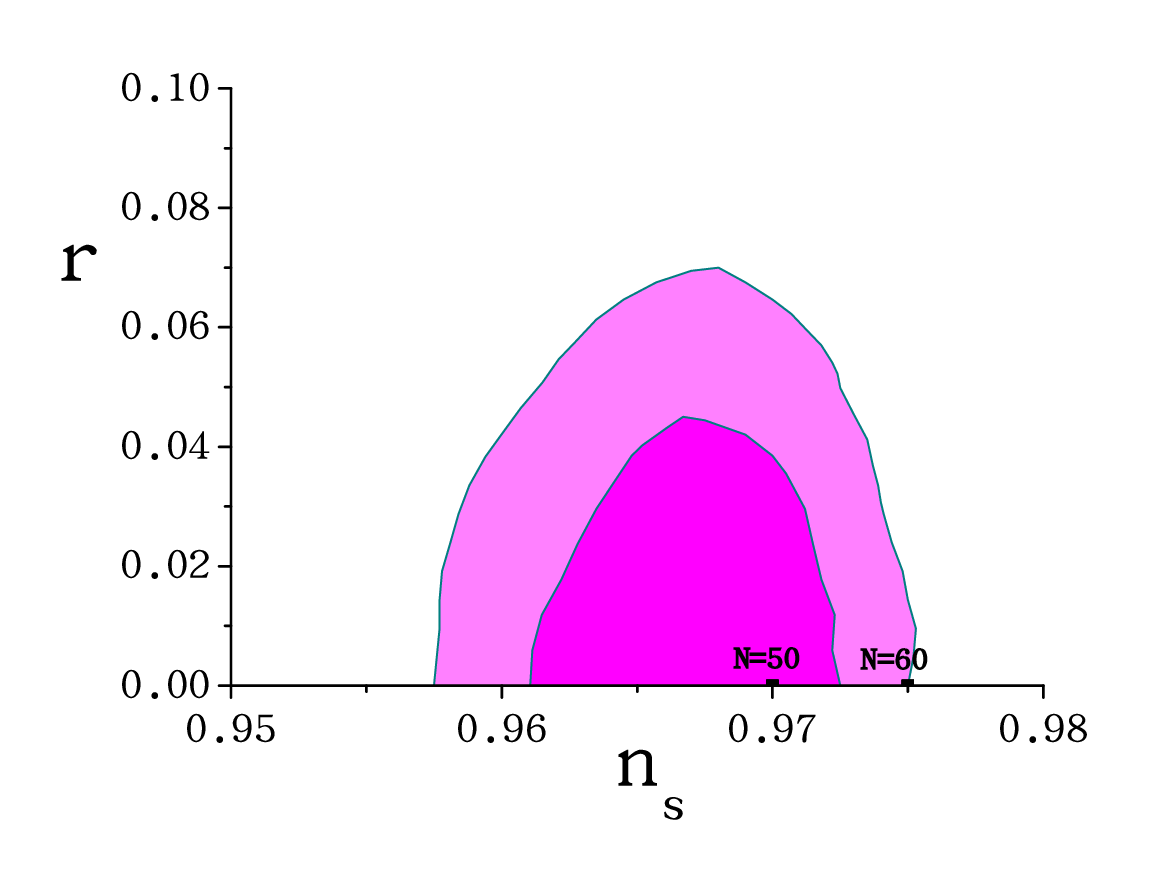}
\caption{{\it{The predictions of Fig. \ref{rns} on top of the 1$\sigma$  and 
2$\sigma$   Planck 2018  TT,TE,EE+lowE+lensing +BK15+BAO results 
\cite{Planck:2018jri} (note that at the scale of the figure all curves 
degenerate at a point very close to the horizontal axis).
}}}
\label{rnsdata}
\end{figure}

In Fig. \ref{rns}   we present the estimated 
tensor-to-scalar ratio versus the spectral index, for three parameter choices. 
In each choice we fix  $\beta  $ and $\gamma$, while  $\alpha$ is the 
parameter that varies along the curve. As we observe, the tensor-to-scalar 
ratio acquires   extra small  values, namely  less than $10^{-5}$. 
{{
The reason behind this feature is the form of the potential (\ref{potential}), 
since at the plateau one 
has   $\varepsilon_V \propto \phi^{-6}$, which leads to    extra 
suppressed tensor-to-scalar 
ratio.}}
Additionally, in  Fig. \ref{rnsdata}  we depict the predictions of Fig. 
\ref{rns} on top of the  Planck 2018    results, where obviously, in the 
scale of the figure all predictions degenerate  at a point very close to the 
horizontal axis.
{Finally, for completeness, let us calculate the running of the spectral 
index $\alpha_\mathrm{s}$ from (\ref{eps333}). Firstly we immediately find that 
  $\alpha_s\simeq -3/(2N_\star^2)$, which leads to 
 $\alpha_s\approx -0.0006$ for   $N_\star=50$ and to 
    $\alpha_s\approx  -0.000417$ for $N_\star=60$. These results are 
consistent at less than $1.3\sigma$ with the ACT DR6+Planck results ($\alpha_s 
= 
+0.0062 \pm 0.0052 $ at $ 1\sigma$ \cite{ACT:2025tim}, which is statistically 
compatible with a 
vanishing running specral index).}

As we can see, the inflationary realization driven by   
potential  (\ref{potential}) with generalized  exponential 
plateau leads to predictions that lie at the core of the  Planck 2018    
results. The extra small tensor-to-scalar ratio, which is the main result of 
the present work,   arises from the steep decay of the slow-roll 
parameter $\varepsilon_V \propto \phi^{-6}$   at large field values. As 
$\alpha/\gamma$ increases $r$ grows slightly but remaining always at extra 
small region, reflecting the 
connection between the modulated plateau and the quadratic minimum. We mention 
that for each $\beta  $ and $\gamma$ choice, $\alpha$ cannot vary arbitrarily 
in order not to spoil the reheating temperature, as we will show in the 
following.
Finally, note that increasing the e-folding number  $N_*$ leads to 
larger $n_{\mathrm{s}}$ and thus values larger than $N_*=60$ are excluded.

These results indicate that the inflationary potential, parameterized as above, 
successfully reproduces the observed properties of the scalar power 
spectrum, both its amplitude and tilt, while predicting an extremely suppressed 
tensor-to-scalar ratio.

 Let us proceed by   examining whether we obtain a successful transition to the 
radiation-dominated era via reheating. As we mentioned above, the quadratic 
minimum of the potential ensures a graceful exit from 
inflation and will induce a reheating phase via coherent oscillations of the 
inflaton field. In 
the framework of minimally-coupled scalar theories the 
reheating temperature $T_{reh}$ is determined by the inflaton decay rate, 
which depends on its mass $m_{\phi}$ and couplings. For the potential 
(\ref{potential}), and for the representative values  $\alpha = \gamma = 1 $, 
$\beta = 1.7\times 10^{-7}$ and $V_0=2  \times 10^{60}\,\text{GeV}^4$ used 
above, the 
inflaton mass near the minimum is:
 \begin{equation}
     m_{\phi} = \sqrt{V''(0)} \approx 2 \times 10^{12}  \text{TeV}.
 \end{equation}
This enhanced mass enables efficient energy transfer through gravitational 
decay, with a decay rate:
\begin{equation}
    \Gamma \sim \frac{m_{\phi}^3}{M_{Pl}^2}\approx 1.38 \times  10^9  
\text{GeV},
\end{equation}
and thus the reheating temperature is derived as  
\begin{equation}
    T_{\text{reh}} \sim \left(\frac{90}{\pi^2g_*}\right)^{1/4} \sqrt{\Gamma 
M_{Pl}} \approx 3.1 \times 10^{16} \,\text{MeV},
\end{equation}
where $g_*\sim 100$ is the effective relativistic degrees of freedom. This 
temperature satisfies the Big Bang Nucleosynthesis (BBN) constraint 
$T_{\text{reh}}\gtrsim 1$MeV, ensuring the synthesis of light elements remains 
unaffected.

Finally, we mention that since the scenario at hand is a single-field model 
and does not include non-canonical kinetic terms, higher-derivative 
interactions, or multi-field dynamics, one     expects suppressed
 non-Gaussianities. 
 
  \subsection{Conformal transformation and $f(R)$ modified gravity}

In the previous subsection we presented a single-field inflationary model based 
on 
the potential (\ref{potential}) with generalized  exponential 
plateau. It would be interesting to perform an inverse conformal 
transformation to the Jordan frame and find the corresponding modified gravity 
$f(R)$ form.
In particular, starting from
\begin{equation}
 \tilde{g}_{\mu\nu} = \Omega^2 g_{\mu\nu}  ,
\end{equation}
with the conformal factor being
\begin{equation}
\Omega^2 = \phi, \quad \text{where}\quad \phi = 
\frac{df(R)}{dR},
\end{equation}
  we can reconstruct $f(R)$ from   $ V(\phi) $ via the 
parametric form 
\begin{eqnarray}
        &&U'(\phi) = R, \\
        &&f(R) = \phi R - U(\phi),
\end{eqnarray}

where 

\begin{equation}
    U(\phi) = \phi^2 V_0 \left[ 1-\exp\left( \frac{-3\alpha 
  \ln^2{\phi}}{2\beta   +3 \gamma   
 \ln^2{\phi}} 
  \right) \right] .
\end{equation}
 
\begin{figure}[ht]
\centering
\hspace{-1.cm}
\includegraphics[scale=.55]{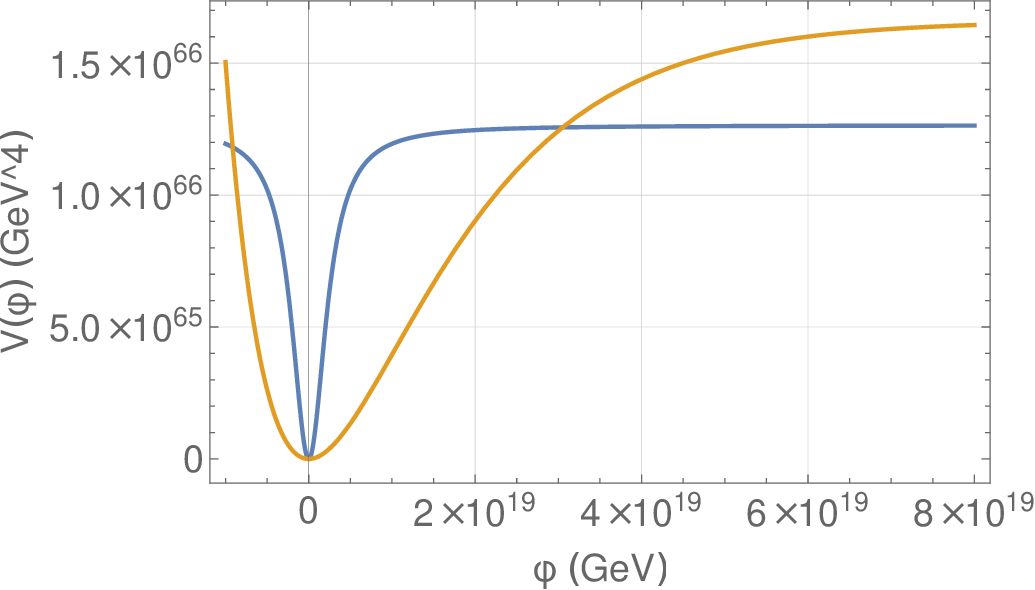}
\caption{{\it{ The  potential  $ V(\phi) = V_0 \left( 1-e^{\frac{-\alpha 
\phi^2}{\beta M_{Pl}^2 +\gamma \phi^2}}\right)$  with generalized  exponential 
plateau, for   $\alpha = \gamma = 1 $, 
$\beta = 1.7 $ and $V_0=2  \times 10^{66}GeV^4$ (blue curve), 
alongside   
Starobinsky 
potential $V_S(\phi)=V_{S0} \left(1 - e^{-\sqrt{\frac{2}{3}} 
\frac{\phi}{M_{\text{Pl}}}} \right)^2
$ for $V_{S0} = 1.66\times 10^{66}GeV^4$ (yellow curve). 
}}}
\label{potentials}
\end{figure}

Unfortunately, for the potential  (\ref{potential}) the above procedure cannot 
be performed analytically since in the last step the function is not 
reversible. However, it can be performed numerically.
As we find, in principle the resulting $f(R)$ can be approximated by a 
polynomial form 
\begin{eqnarray}
 f(R)=\sum_i \zeta_i R^i,
\end{eqnarray}
where the  absolute 
values of the coefficients $\zeta_i$ tend to decrease      after the first 
terms. 
In the sample case where  $\alpha = 1$, $\beta  = 1$, 
$\gamma = 1$ and $V_0 = 2$, in $M_{\text{Pl}}$ units, we find that 
$\zeta_0\approx 0.04$
$\zeta_1\approx 0.54$, 
$\zeta_2\approx 0.078$,
 $\zeta_3\approx 0.00017$,
 $\zeta_4\approx  {\mathcal{O}}
 (10^{-7})$, etc.
 However, in the physically interesting cases examined in 
the previous sections, where $\beta = 1.7\times 10^{-7}$ and $V_0=2  \times 
10^{60}\,\text{GeV}^4$, with $\alpha$ and $\gamma$ of the order of 
$ 1 $, we find 
that one needs to keep many terms in the expansion, with changing signs by 
turn, which is an indication that the power-law-expansion is not very efficient.
Nevertheless, it is exactly these non-zero corrections to 
the well-behaved Starobinsky model, that improve the behavior of the 
tensor-to-scalar ratio and suppress it to extra small values. 
 For completeness, in Fig. \ref{potentials} we depict 
potential (\ref{potential}) with generalized  exponential 
plateau,  alongside   
Starobinsky 
potential  $V_S(\phi)=V_{S0} \left(1 - e^{-\sqrt{\frac{2}{3}} 
\frac{\phi}{M_{\text{Pl}}}} \right)^2
$,  in order for the differences of the two potentials to be 
more transparent.

\section{Conclusions}

Inflation is a necessary ingredient of the concordance model of cosmology, 
since it offers a successful solution to the  horizon and flatness problems, as 
well as it generates the perturbation structure that is visible in the cosmic 
microwave background (CMB) radiation and subsequently in the large scale 
structure. Nevertheless, obtaining a sufficiently small tensor-to-scalar ratio 
in order to be compatible with the stringent  observational bounds, proves to 
be a challenging task. In particular, since the traditional  large-field models 
based on simple monomial potentials typically predict  significant tensor 
amplitudes, one should try to construct various extended inflationary models 
with plateau-like potentials, obtained from various scalar-tensor models or 
from modified gravity.

In this work we considered a standard minimally-coupled scalar-field 
inflationary scenario, but which is based on a new  potential, 
with suitably generalized plateau features. In particular, we consider a 
specific three-parameter potential (and the potential amplitude as an extra 
parameter) given in (\ref{potential}), which has a flatter plateau and a 
steeper well compared to the Starobinsky potential in the Einstein frame, that 
leads to  extra small tensor-to-scalar ratios compared to Starobinsky 
inflation.

Investigating the inflationary realization we showed that the generalized 
plateau can guarantee a prolonged period of slow-roll 
inflation and a successful exit. Additionally, the steeper minimum compared to  
other inflationary models, such as the Starobinsky one, leads to significantly  
suppressed tensor perturbations, and thus to an extra-small tensor-to-scalar 
ratio $r$. In particular, fixing the potential amplitude $V_0$ and the 
dimensionless parameter $\beta $ in order to obtain a  scalar power 
spectrum 
amplitude $  A_s\approx 2 \times 10^{-9}  $
and spectral tilt $n_s\approx 0.97$, we varied  the remaining two 
parameters, namely $\alpha$ and $\gamma$, in order to obtain a suppressed
 tensor-to-scalar 
ratio.

As we showed, we were able to obtain $r$ values  less than $10^{-5}$, 
well within the  Planck 2018    results. The reason for this feature is 
the steep decay of the slow-roll parameter $\varepsilon_V \propto \phi^{-6}$   
at large field values, since as $\alpha/\gamma$ increases then $r$ grows 
slightly but remaining always at extra 
small region, reflecting the connection between the modulated plateau and the 
quadratic minimum. Moreover, we calculated the reheating temperature showing 
that in order to be in agreement with observations the parameter $\alpha$ 
should remain within specific bounds.
Finally, performing a conformal transformation to the Jordan frame 
we showed that the considered potential corresponds to higher-order corrections 
to Starobinsky potential in the Einstein frame, and these corrections are the 
reason for the improved behavior of the tensor-to-scalar ratio.

In summary, the proposed inflationary scenario based on a new potential with a 
generalized plateau, successfully reproduces the observed properties of the 
scalar power spectrum, both its amplitude and tilt, while predicting an 
extremely suppressed tensor-to-scalar ratio, and the fact these features are 
obtained within the simple, minimally-coupled  scalar-field models, is an 
additional advantage. One should try to provide a theoretical justification for 
the specific potential form, or try to obtain it within other modified gravity 
frameworks beyond the standard curvature one, such as torsional or 
non-metricity theories of gravity 
\cite{Kouniatalis:2024gnr,Capozziello:2022tvv}. Such a detailed investigation 
and/or model 
reconstruction lies beyond the scope of this first analysis on the subject, and 
it is left for a future project.

\begin{acknowledgments} 
We would like to  thank Theodoros Papanikolaou for   insightful discussions.
The authors acknowledge the contribution of the LISA CosWG, and of   COST 
Actions   CA21136 ``Addressing observational tensions in cosmology with 
systematics and fundamental physics (CosmoVerse)'', CA21106 ``COSMIC WISPers 
in the Dark Universe: Theory, astrophysics and experiments'', 
and CA23130 ``Bridging high and low energies in search of quantum gravity 
(BridgeQG)''.

\end{acknowledgments}


\end{document}